\documentclass[a4paper,twoside]{article}

\usepackage{epsfig}
\usepackage{subcaption}
\usepackage{calc}
\usepackage{amssymb}
\usepackage{amstext}
\usepackage{amsmath}
\usepackage{amsthm}

\usepackage{multicol}
\usepackage{pslatex}
\usepackage{apalike}
\usepackage{algorithm2e}
\usepackage[bottom]{footmisc}
\usepackage{comment}
\usepackage{url}
\usepackage{blindtext}
\usepackage{hyperref}
\usepackage{nicefrac}
\usepackage{braket}
\usepackage{mathtools}
\usepackage{bbold}
\usepackage[T1]{fontenc}
\usepackage{orcidlink}

\DeclareMathOperator*{\argmax}{arg\,max}
\DeclareMathOperator*{\argmin}{arg\,min}

\makeatletter
\def\blfootnote{\xdef\@thefnmark{}\@footnotetext}
\makeatother

\usepackage{SCITEPRESS}     

\begin{document}

\title{Introducing Reduced-Width QNNs, an AI-inspired Ansatz Design Pattern}

\author{\authorname{Jonas Stein\sup{1,2,\text{\textdagger}}\orcidlink{0000-0001-5727-9151}, Tobias Rohe\sup{1,\text{\textdagger}}, Francesco Nappi\sup{1}, Julian Hager\sup{1}, David Bucher\sup{2}\orcidlink{0009-0002-0764-9606},\\ Maximilian Zorn\sup{1}\orcidlink{0009-0006-2750-7495}, Michael Kölle\sup{1}\orcidlink{0000-0002-8472-9944} and Claudia Linnhoff-Popien\sup{1}\orcidlink{0000-0001-6284-9286}}
\affiliation{\sup{1}Mobile and Distributed Systems Group, LMU Munich, Germany}
\affiliation{\sup{2}Aqarios GmbH, Germany}
\email{jonas.stein@ifi.lmu.de}
}

\keywords{Quantum Computing, Variational Quantum Circuits, Circuit Optimization, Variational Quantum Eigensolver}

\abstract{Variational Quantum Algorithms are one of the most promising candidates to yield the first industrially relevant quantum advantage. Being capable of arbitrary function approximation, they are often referred to as Quantum Neural Networks (QNNs) when being used in analog settings as classical Artificial Neural Networks (ANNs). Similar to the early stages of classical machine learning, known schemes for efficient architectures of these networks are scarce. Exploring beyond existing design patterns, we propose a \emph{reduced-width} circuit ansatz design, which is motivated by recent results gained in the analysis of dropout regularization in QNNs. More precisely, this exploits the insight, that the gates of overparameterized QNNs can be pruned substantially until their expressibility decreases. The results of our case study show, that the proposed design pattern can significantly reduce training time while maintaining the same result quality as the standard "full-width" design in the presence of noise.} 

\onecolumn \maketitle \normalsize \setcounter{footnote}{0} \vfill

\section{\uppercase{Introduction}}
\label{sec:introduction}
\blfootnote{\textsuperscript{\textdagger}These authors contributed equally.}
A central goal of current quantum computing research is achieving an industrially relevant quantum advantage. As immense computational requirements of known quantum algorithms enabling provable speedups exceed prevailing hardware capabilities in the present NISQ era~\cite{preskill2018quantum}, less demanding approaches like Quantum Neural Networks (QNNs) are of great interest, especially as they have already shown very promising results towards a possible quantum advantage~\cite{Abbas2021}. QNNs closely resemble classical Artificial Neural Networks (ANNs), as they are composed of concatenated, parameterized functions capable of approximating arbitrary functions~\cite{PhysRevA.98.032309}. The core difference between QNNs and ANNs is the dimensionality of the underlying vector space, which is exponentially larger in the former with respect to the space requirements in qubits/neurons~\cite{PhysRevA.98.032309}. 

A key advantage of QNNs compared to established quantum algorithms like those of \cite{365700} or \cite{10.1145/237814.237866} is their flexibility towards respecting restricted hardware capabilities. More precisely, QNNs can be shaped to fit a maximal width and depth of the circuit to be executed on the QPU and can thus minimize errors caused, e.g., by decoherence. Analogous to the historical progress on ANNs, where many useful and performant network architectures and design patterns emerged, recent literature on QNNs explores productive quantum circuit design patterns \cite{Cerezo2021,https://doi.org/10.1002/qute.201900070,PhysRevA.103.032430,a12020034,Havlicek2019}. Similar to ANNs, these techniques are used to find a sufficiently good trade-off between expressibility and trainability \cite{https://doi.org/10.1002/qute.201900070}. 

In classical machine learning, this problem is predominantly solved by utilizing heavily overparameterized models and \emph{regularization} techniques to facilitate efficient trainability. A particularly powerful regularization technique is dropout, in which parts of the model are randomly deleted for each training step. Among other effects, this introduces noise to the network architecture and therefore hinders overfitting~\cite{srivastava14a}. Meanwhile, even though research on regularization techniques for QNNs is still in its early stages, dropout has also shown first promising results~\cite{scala2023general,kobayashi2022overfitting}. Analyses on instances of overparameterized QNNs in~\cite{scala2023general} show no reduction of expressibility in the presence of substantial gate pruning. In this article, we use these insights as inspiration for a circuit architecture design pattern, that directly employs a gate-pruned version of overparameterized QNNs. This is explicitly not targeted towards substituting dropout, but meant as an exploration of the necessity of the standard full-width layer design in the first place. 

To test the effectiveness of our approach, we evaluate QNNs of a fixed circuit architecture before and after pruning. Our main contributions in this paper can be summarized as the following:
\begin{itemize}
    \item We propose a novel, AI-inspired QNN design pattern, that can notably also be applied as a pruning technique for existent QNNs.
    \item We conduct an empirical case study of this design pattern for the maximum cut optimization problem in a noisy environment.
\end{itemize}

This article is structured as follows. Sec.~\ref{sec:background} outlines necessary background knowledge on solving optimization problems with QNNs. Sec.~\ref{sec:relatedwork} gives an overview of known ansatz design patterns from literature. Subsequently, we present our novel design pattern in Sec.~\ref{sec:methodology}. Following this, Sec.~\ref{sec:evaluation} displays a case study evaluation in the presence of noise. Finally, Sec.~\ref{sec:conclusion} concludes our findings and motivates possible future work.

\section{\uppercase{Background}}
\label{sec:background}
In the following, we describe the core algorithms and techniques employed in our methodology. We show how QNNs can be utilized to solve specific optimization problems like the maximum cut problem as part of the Varational Quantum Eigensolver (VQE). 

\subsection{Quantum Neural Networks}\label{subsec:QNNs}
\begin{figure}
    \centering
    \includegraphics[width=0.3\textwidth]{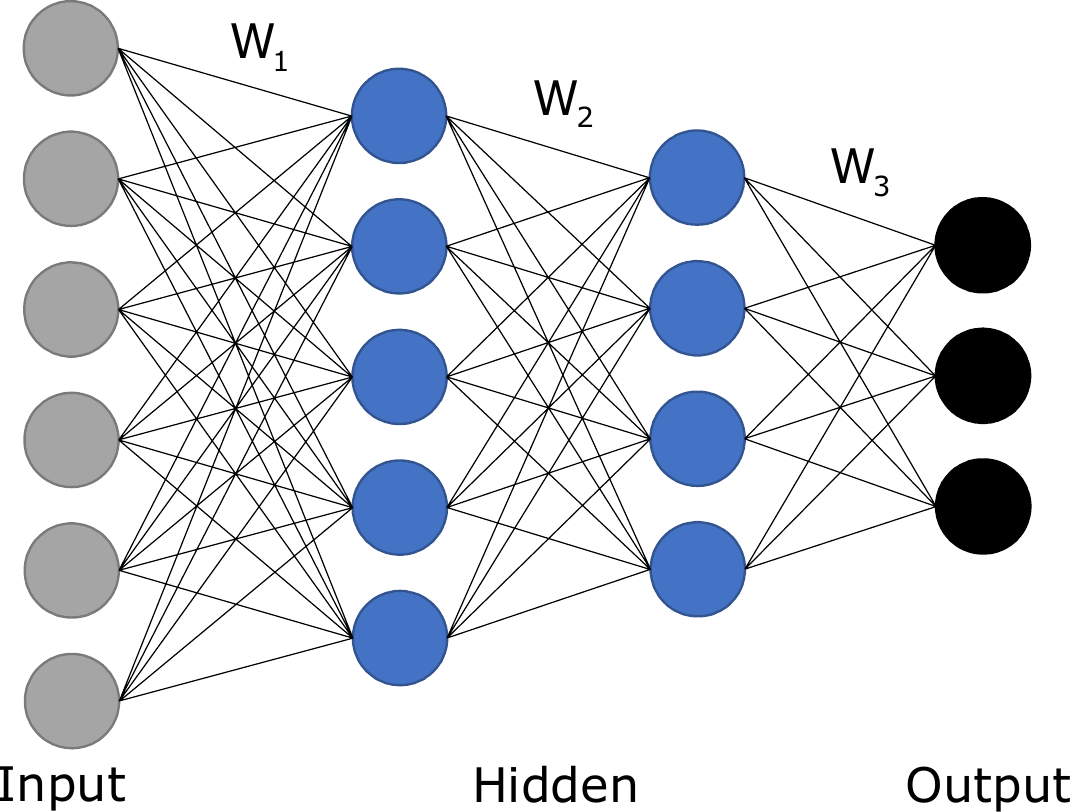}
    \caption{Example network architecture of the encoder in an autoencoder.}
    \label{fig:autoencoder-architecture}
\end{figure}

In current literature, the term Quantum Neural Network is predominantly used to describe a Parameterized Quantum Circuit (PQC) utilized for tasks typical for classical Artificial Neural Networks~\cite{Beer2020,mcclean2018barren}. Although the terminology suggests a close similarity between QNNs and ANNs, there exist significant differences. A core distinction is that QNNs do not consist of individual neurons, instead they are comprised of qubits, whose states can be changed using (parameterized) quantum gates, that can involve interactions with other qubits. Despite the substantial differences between ANNs and QNNs, there also exist strong commonalities, foremost the shared capability of universal function approximation, i.e., ANNs and QNNs can both be described as parameterized functions, that can approximate arbitrary functions under certain conditions~\cite{PhysRevA.103.032430}. Moreover, by representing QNNs and feedforward ANNs as mathematical functions, we can observe clear similarities. A feedforward neural network is mathematically represented by concatenations of typically non-linear activation functions for all neurons in each layer~\cite{SCHMIDHUBER201585}. In contrast, QNNs typically model concatenations of linear layers with a single non-linear layer at the end, which is responsible for the measurement, i.e., generating a classical output~\cite{PhysRevA.98.032309}. For examples of feedforward ANNs and QNNs, see Figs.~\ref{fig:autoencoder-architecture} and~\ref{fig:basicCircuit}.

Mathematically, the final state of a QNN composed of $n \in \mathbb{N}$ qubits can be formalized as denoted in Eq.~\ref{eq:QNN-result}\footnote{This formalization explicitly disregards intermediate measurements, in-line with predominant literature.}. In this formalization, we started in the standard initial state $\ket{0}^{\otimes n}$ and subsequently applied a parameterized unitary operation $U\left(\theta\right)$, with parameters $\theta\in\mathbb{R}^m$, where $m\in \mathbb{N}$ denotes the number of parameters. In practice, $U$ is predominantly comprised of parameterized single-qubit and two-qubit gates to match native gate sets of existing quantum computers. 
\begin{align}
\label{eq:QNN-result}
    \ket{\psi\left(\theta\right)}\coloneqq U\left(\theta\right)\ket{0}^{\otimes n}
\end{align}

To extract classical information from the final state of the QNN, we typically measure each qubit $\ket{\psi_i}$ in the computational basis $\left\lbrace\ket{0},\ket{1}\right\rbrace$ using the hermitian measurement operator $M=I^{i-1}\otimes\sigma_z\otimes I^{n-i}$, where $\sigma_z$ denotes the Pauli-$Z$ operator. These measurements yield a probability distribution over bitstrings $\ket{x}\in\mathbb{C}^{2^n}$ representing the output of the quantum circuit, where $\ket{x}$ occurs at probability $\left|\braket{x\mid\psi\left(\theta\right)}\right|^2$.


Having presented an overview of the circuit components of a QNN, we now address parameter training approaches. Analogous to ANNs, two main types of parameter optimization approaches exist: gradient-based and derivative-free methods~\cite{PhysRevA.98.032309}. While QNNs can be trained with essentially the same techniques as ANNs, i.e., using approaches like gradient descent, genetic algorithms, and others, known gradient computations for QNNs are fundamentally more expensive. Using backpropagation, the gradient in feedforward ANNs can be calculated as fast as a forward pass of the ANN, whereas all known quantum gradient computations need $\mathcal{O}\left(m\right)$ executions of the QNN, where $m\in \mathbb{N}$ denotes the number of parameters~\cite{PhysRevA.98.032309,PhysRevA.99.032331}. The standard method to calculate the gradient of a parameterized quantum circuit is the parameter shift rule, which is used to calculate each partial derivative $\nicefrac{\partial}{\partial \theta_i} U\left(\theta\right)\ket{0}^{\otimes n}$ based on Eq.~\ref{eq:parametershift}, given that all parameters are enclosed in single qubit gates and $\theta^\pm\coloneqq \left(\theta_1,...,\theta_{i-1},\theta_i\pm\nicefrac{\pi}{2},\theta_{i+1},...,\theta_m\right)$~\cite{PhysRevA.98.032309,PhysRevA.99.032331}.
\begin{align} \label{eq:parametershift}
    \dfrac{\partial}{\partial \theta_i} U\left(\theta\right)\ket{0}^{\otimes n} = \dfrac{U\left(\theta^+\right)\ket{0}^{\otimes n} - U\left(\theta^-\right)\ket{0}^{\otimes n}}{2}
\end{align}
Given this linear runtime increase in the gradient computation for QNNs, achieving better solution quality or a quick convergence to sufficiently good parameters is essential for QNNs.

\subsection{The Maximum Cut Problem in Quantum Optimization}\label{subsec:maxcut}
The \emph{maximum cut} problem, is an NP-hard combinatorial optimization problem with the objective of partitioning the nodes $v_i\in V$ of a given graph $G\left(V,E\right)$ into two complementary sets $A\dot\cup B=V$ such that number of edges between $A$ and $B$ is maximal. Encoding a partitioning via a binary vector $s\in\left\lbrace -1,1\right\rbrace^n$, where $n\coloneqq \left|V\right|$ and $s_i=1$ $\Leftrightarrow$ $v_i\in A$ and $s_i=-1$ $\Leftrightarrow$ $v_i\in B$, the search for the optimal partition can be formulated as shown in Eq.~\ref{eq:maxcutobjfun}~\cite{10.3389/fphy.2014.00005}. This formulation naturally evolves after observing that $\nicefrac{\left(1-s_i s_j\right)}{2}\in\left\lbrace 0,1\right\rbrace$ equals one iff the edge $\left(v_i,v_j\right)\in E$ connects $A$ and $B$.
\begin{align}\label{eq:maxcutobjfun}
    &\argmax_{s\in\left\lbrace -1,1\right\rbrace^n}\sum_{(v_i,v_j)\in E} \dfrac{1-s_i s_j}{2}\nonumber\\
    =&\argmin_{s\in\left\lbrace -1,1\right\rbrace^n}\sum_{(v_i,v_j)\in E} s_i s_j
\end{align}
We have chosen this $\pm 1$ encoding of the solution, as it can directly be transformed into a quantum Hamiltonian $\hat{H}$ that represents the solution space of this optimization problem, which is the standard approach in quantum optimization exploiting that Ising Hamiltonians are isomorphic to the NP-hard QUBO. For this, we merely need to substitute $s_i$ using the map $s_i\mapsto \sigma^z _i \coloneqq  I^{\otimes i-1} \otimes \sigma_z \otimes I^{\otimes n-i}$, where $I$ denotes the two dimensional identity matrix. Therefore, we can define $\hat{H}\coloneqq \sum_{\left(v_i,v_j\right)\in E}\sigma^z _i \sigma^z _j$.

Switching to a $0/1$ encoding, i.e., $x\in\left\lbrace 0,1\right\rbrace^n$ with $x_i=0 \Leftrightarrow v_i\in A$ and $x_i=1 \Leftrightarrow v_i\in B$, $\hat{H}$ can be understood as a mapping between solutions $x$ and the value of the objective function. More specifically, its eigenvalues for a given eigenvector $\ket{x}$ are the values of the objective function, where $x$ represents the $0/1$ encoding of the solution.

\subsection{The Variational Quantum Eigensolver}
The Variational Quantum Eigensolver (VQE) is an algorithm targeted towards finding the ground state of a given Hamiltonian $\hat{H}$ using the \emph{variational method}~\cite{Peruzzo2014}. The term \emph{variational method} describes the process of making small changes to a function $f$ to find minima or maxima of a given function $g$~\cite{lanczos2012variational}. In the case of the VQE, $f:\theta \mapsto U\left(\theta\right)\ket{0}^{\otimes n}$ describes a parameterized quantum circuit yielding the state $\ket{\psi\left(\theta\right)}\coloneqq U\left(\theta\right)\ket{0}^{\otimes n}$  which represents a guess of the $\argmin$ of the function $g:\ket{\varphi}\mapsto \bra{\varphi} \hat{H} \ket{\varphi}$. More specifically, the VQE uses parameter training techniques on the given ansatz $U\left(\theta\right)$ to generate an output state $\ket{\psi\left(\theta\right)}$ that approximates the ground state of $\hat{H}$.

For our purposes, i.e., solving an optimization problem, we need to (1) define a Hamiltonian encoding our objective function and (2) select a suitable ansatz in order to use the VQE algorithm. While we can use the procedure exemplified in section~\ref{subsec:maxcut} to define a Hamiltonian $\hat{H}$, the selection of the ansatz and thus a circuit architecture is not thoroughly understood yet. Although literature on automatized quantum architecture search tools is emerging~\cite{Du2022}, most ansätze are designed based on empiric experience~\cite{sim2019expressibility}. For more details on ansatz design, see Sec.~\ref{sec:relatedwork}.



\section{\uppercase{Related Work}}
\label{sec:relatedwork}

A suitable circuit design in QNNs is essential for achieving good results in variational quantum computing and is actively researched in literature~\cite{sim2019expressibility}. While some design patterns like a layerwise circuit structure with alternating single qubit rotations and entanglement gates, have become standard practice~\cite{sim2019expressibility}, the list of established circuit elements still is considerably smaller than that in classical neural networks. A central problem when trying to design ansätze is the trade-off between expressibility and trainablity. Reasonably complex models demand for expressive circuits, which in turn can quickly lead to the problem of \emph{barren plateaus}, a trainability issue where the probability that the optimizer-gradient becomes exponentially small is directly dependent on the number of qubits~\cite{mcclean2018barren}. Moreover, barren plateaus are found to be dependent on the choice of the cost function itself \cite{cerezo2021variational} as well as the number of parameters to optimize simultaneously~\cite{skolik2021layerwise,holmes2022connecting}.


Classical deep learning faced a similar problem in \emph{vanishing gradients}, e.g., for recurrent neural networks \cite{hochreiter1998vanishing}, which eventually lead to the adoption of more specialized network architectures like the long short-term memory (LSTM) networks \cite{hochreiter1997long} for sequence processing. Similarly, the current trend of improving variational architectures focuses on optimizing the designs of quantum circuit ansätze, e.g., via optimization procedures starting from randomized circuits \cite{fosel2021quantum}. In this context, our work shows that pruning gates of the standard full-width circuit architecture can substantially increase trainability while retaining the same solution quality. 

\section{\uppercase{Methodology}}
\label{sec:methodology}
In the subsequent sections, we describe our reduced-width ansatz design pattern in two different variants as pruned versions of a baseline ansatz. Concluding our methodology, we describe our approach to parameter training and problem instance generation.

\subsection{Full-width Design}\label{subsec:ansatz}
In the following, we introduce the design of our baseline ansatz, which we coin \emph{full-width design}. We choose a very simple circuit architecture consisting of \texttt{BasicEntanglerLayers}\footnote{For details and the implementation, see \url{https://docs.pennylane.ai/en/stable/code/api/pennylane.BasicEntanglerLayers.html}.}, as it resembles the most basic circuit architecture and is ubiquitously used for constructing variational quantum circuits in quantum machine learning.

\begin{figure*}[!ht]
  \centering
  \includegraphics[width=\textwidth]{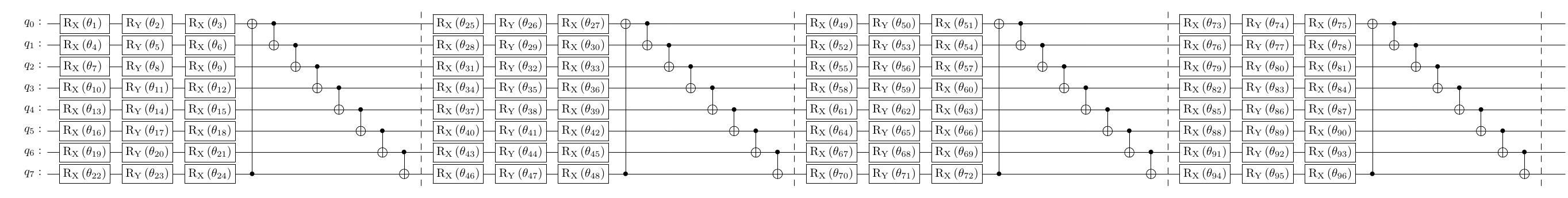}
  \caption{The full-width design in which gates are applied over the full width of the circuit throughout all layers. The circuit is displayed with four layers, in-line with its configuration in our evaluation.}
  \label{fig:basicCircuit}
\end{figure*}

More specifically, this baseline circuit is constructed of two components. The first is a sequence of parameterized gates, i.e., a Pauli-$X$, a Pauli-$Z$ and another Pauli-$X$ rotation, applied to each qubit, allowing for the implementation of arbitrary single qubit operations. The second is a circular CNOT-entanglement, i.e., CNOTs applied to all neighboring qubit pairs, using the maximum width of the circuit. We combine these single- and two-qubit gate components into a single layer, where every circuit is composed of four such layers to analyse scaling behavior. The entire circuit with all its layers is illustrated in Fig.~\ref{fig:basicCircuit}. In the following sections, we will refer to this circuit design as the \textit{full-width circuit} or \textit{full-width design}, as it uses the same maximum width throughout all layers. All the circuits utilized in this study are derived from this full-width circuit, where variations are introduced through the removal of gates.

\begin{figure*}[!ht]
  \centering
  \includegraphics[width=\textwidth]{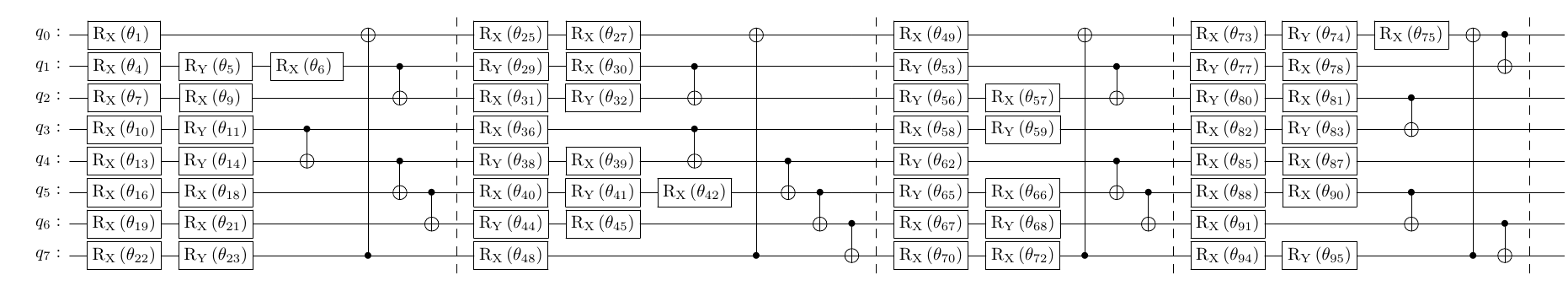}
  \caption{The random-width design in which a specific number of gates are deleted from the full-width design at random positions. The number of gates removed gates equals the number of removed gates in the reducing-width circuit. The circuit is displayed with four layers, in-line with its configuration in our evaluation.}
  \label{fig:randomCircuit}
\end{figure*}

\begin{figure*}[!ht]
  \centering
  \includegraphics[width=\textwidth]{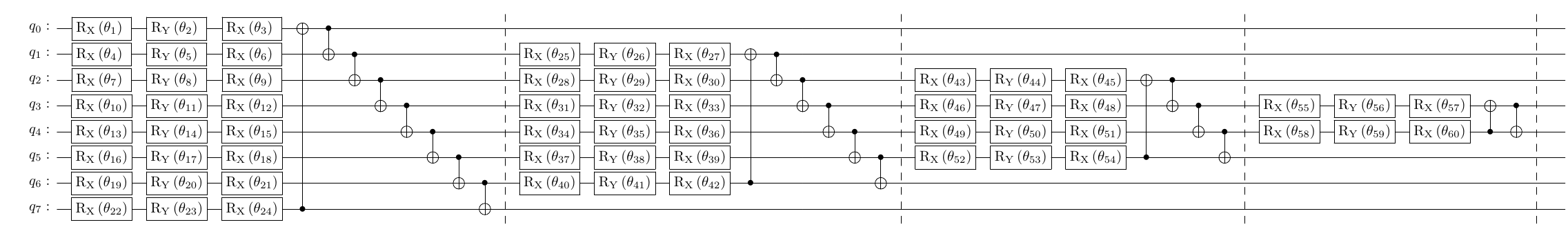}
  \caption{The reducing-width design in which the width is reduced from layer to layer by a magnitude of two. With this design approach, we prune the circuit to a version which uses much fewer gates. The circuit is displayed with four layers, in-line with its configuration in our evaluation.}
  \label{fig:reducingCircuit}
\end{figure*}

\subsection{Reduced-width Design}\label{subsec:pruning}
In this section, we introduce two different approaches that reduce the width of a given ansatz by pruning gates in a layerwise manner. The first approach works analog to the dropout employed in \cite{scala2023general} and randomly prunes the a fixed number of gates from the circuit (for reference, see  Fig.~\ref{fig:randomCircuit}). To test how the structure of the employed pruning affects the performance, we propose a second approach, which gradually increases the number of pruned gates for subsequent layers (for reference, see  Fig.~\ref{fig:reducingCircuit}).

Starting with the detailed discussion of the second approach, we review a key result of \cite{scala2023general}, which states that the deeper the circuits, the more gates can be pruned before effectively reducing the expressibility. This is very sensible, as their results on the effectiveness of pruning are conditioned on present overparameterization, which only occurs for reasonably deep QNNs. Furthermore, we argue, when entanglement between the qubits is sufficiently high (i.e., in general after some ansatz layers), every gate executed on a subset of these qubits intrinsically also acts on all qubits, as they must be regarded as one system. A well-known example in which exactly this happens, is the ancilla quantum encoding as part of the quantum algorithm for solving linear systems of equations~\cite{PhysRevLett.103.150502,stougiannidis2023approximative}, in which controlled rotations on a single ancilla qubit substantially change the state of an entire qubit register.


In our implementation, we reduce the circuit width in every layer by a magnitude of two in a symmetric manner. This implies that the outermost qubits manipulated in the previous layer are no longer altered directly in the appended layer. This reduction applies to both single-qubit gates and two-qubit gates. Consequently, the circular entanglement is applied solely to the remaining inner qubits, so that gates are not only deleted, but also reconnected. A visualization of our reducing-width ansatz can be found in Fig.~\ref{fig:reducingCircuit}.

The number of reduced single- and two-qubits gates is dependent on the number of qubits $n$, the number of layers $l$, and the magnitude by which the circuit width is reduced per layer $m$. Further, we have to distinguish between the reduction in one-qubit gates $\sum_{i=1}^{l} (i-1)3m$, and the reduction in two-qubit gates $\sum_{i=1}^{l} (i-1)m$. The proposed reduction is naturally restricted by the number of qubits in the circuit, i.e., $(l-1)m < n$ must be satisfied, so that the reduction can be conducted when starting from the full-width circuit.

In our case (i.e., using eight qubits, four ansatz layers, and a reduction of the circuit width per layer of two in magnitude), we reduce the number of one-qubit gates by 36, and the number of two-qubit gates by 12. Respectively, this corresponds to a reduction of one-qubit gates by 37.5\%, and a 50\% reduction for two-qubit gates. Beside the hypothesized positive effects that a reduction of the number of gates has on the parameter optimization process, the proposed circuit design is naturally also more noise tolerant. More specifically, the reduced number of gates and consequently the lower circuit depth reduce the total noise accumulated from imperfect gate fidelities and decoherence \cite{Hashim2023}. We therefore expect our approach to be the most beneficial in a noisy environment.


As stated in the beginning of this section, we also introduce the so called random-width design. Here, we prune the same number and the same kind of gates from the full-width design as in the reducing-width approach. However, instead of removing these gates in a systematic way, as done in the reducing-width design, we randomly determine the gates for removal. Therefore, the number and kind of gates compared to the reducing-width design stays the same, but the sequence and position of the gates differ. An exemplary circuit resulting from the random-width design is shown in Fig.~\ref{fig:randomCircuit}. To achieve statistical relevance in the structure of the pruned gates, the results of three different prunings are averaged over each problem instance in the evaluation.

\subsection{Training Method}\label{subsec:training}
Aiming for a test case in which our approach can show its full potential (see Sec.~\ref{subsec:pruning}), we decide to restrict our evaluation to a noisy environment. Therefore, we employ layerwise learning, which is a method specifically targeted at coping with noise during the training and has been shown to increase result quality in many cases significantly \cite{skolik2021layerwise}. A small prestudy also shows that this is the case for our setup. The training conceptually starts with the first layer of the respective circuit individually, i.e., without any other layers present in the circuit. Subsequently, the parameterized gates get initialized with zeroes to approximate an identity layer similar to the approach proposed in~\cite{liu2022layer} and the parameter training is started. For the training process, we use the COBYLA optimizer \cite{powell1994cobyla}, where the maximum number of iterations is set to 300 to allow for the learning curve to converge sufficiently, according to empirical pretests conducted for this study. After the training of the first layer is completed, the trained parameters get saved, and a new quantum circuit is constructed. The new circuit now implements two layers and re-uses the previously trained and saved parameters for the initialization of the first layer. The parameters of the gates added in the next layer of the circuit are initialized with zero, while the subsequent training-process only trains the newly added layer of the respective circuit. This training method now repeats itself until all four layers are implemented and trained. This training approach therefore aims at improving on every previously achieved solution iteratively, similar to techniques like reverse annealing. 

\subsection{Problem Instance Generation}\label{subsec:problem}
To evaluate the performance of the three different design approaches, we solve the maximum cut problem for 50 randomly generated 8-node Erdős-Rényi graphs~\cite{Erdos2022OnRG} using a VQE setup with the respective circuit design as the ansatz. These graphs are generated in a random manner, where the number of nodes, as well as the probability that an edges exists between each pair of nodes, are given as input. We use eight nodes to match the qubit count and choose an edge probability of 50\% to yield reasonably realistic problem instances, analogous to~\cite{9459509}. 

\section{\uppercase{Evaluation}}
\label{sec:evaluation}
Our experiments are conducted with the three different ansätze shown in Figs.~\ref{fig:basicCircuit}, ~\ref{fig:reducingCircuit} and \ref{fig:randomCircuit} aiming to solve 50 maximum cut problem instances using a VQE approach and layerwise learning. As motivated in Sec.~\ref{subsec:pruning}, we employ depolarizing noise effects in our circuit simulations: The one-qubit and two-qubit gate error rate were both set to 1\%, which roughly coincides with the gate fidelities of recent IBM QPUs. To facilitate these comparatively compute intensive circuit simulations, we run our experiments on an Atos Quantum Learning Machine (QLM). 

In the following two sections, we investigate the runtime and result quality of the proposed reduced-width ansatz design pattern. To facilitate a basic scaling analysis, we run all experiments with an iteratively increasing number of maximal layers up to four total.

\subsection{Execution Times}\label{subsec:execution_times} 

\begin{figure}[t]
  \centering
  \includegraphics[width=\columnwidth]{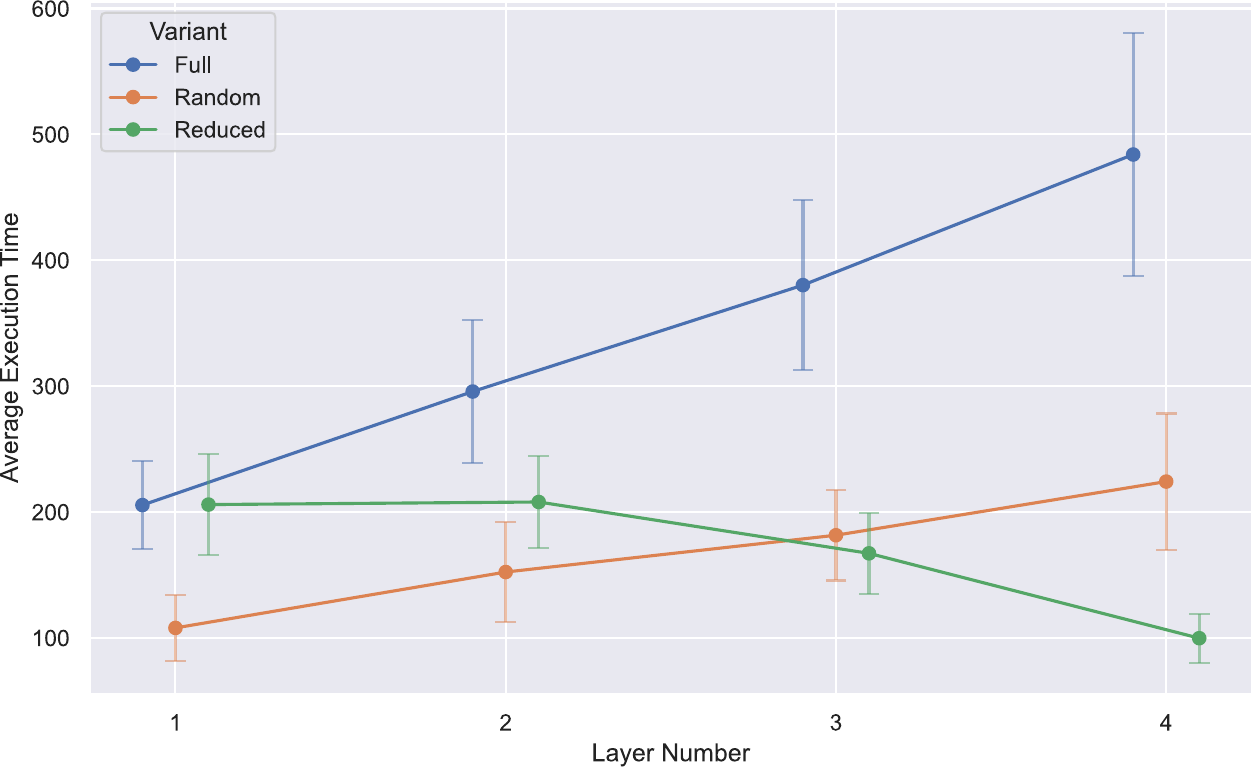}
  \caption{Average execution time necessary to train each layer for the different circuit design schemes.}
  \label{fig:average_time}
\end{figure}

We measure the time complexity in terms of the Average Execution Time (AET), which is comprised of the overall circuit execution and the time spent on training. For all circuit-designs and their different number of layers, the AET is measured in seconds over the 50 synthetically generated problem instances. The results are displayed in Fig.~\ref{fig:average_time}. 

For the full-width design and the random-width design, we observe increasing AETs with increasing number of layers. This effect is stronger for the full-width design than for the random-width design. We explain those results by the two major tasks the simulator is performing: The circuit simulation and the parameter optimization. With increasing number of layers, the number of gates to simulate increases, which in turn leads to increasing circuit simulation times. However, the training time dominates the circuit execution time heavily as a closer inspection of the data shows (e.g., in layer one). Since the random-width design implements fewer gates overall and as the reduction of gates and parameters is distributed randomly across all layers, the AET curve has a flatter slope than the curve of the full-width design. 

When analyzing the AET development of the reducing-width approach, we observe another pattern. In layer one, the AET coincides with that of the full-width design, as both designs implement the same circuit in the first layer. On the other hand, the reducing-width approach has a lower AET here, as its pruning already starts in layer one. From layer one to layer two, the AETs slightly increase for the reducing-width design, while for the remaining layers we observe significant decreases of its AETs. We can also explain this behavior by looking at the supposed development of the two main components of the AET. We assume, that the circuit simulation time increases throughout the layers at a decreasing pace, however, the absolute parameter optimization time decreases faster, as fewer gates are appended to the circuit and therefore fewer parameters need to be trained. While the effect of increasing circuit simulation time is still stronger in the step from the first layer to the second layer than the effect of decreasing optimisation time, this ratio changes constantly, starting at the transition from layer two to three. The AET is almost halved from layer one to layer four. These results indicate good scalability properties of the reducing-width design in regards to training time.

\subsection{Average Approximation Ratios}\label{subsec:approx_ratios}


In order to assess the performance of different design schemes qualitatively, we investigate their Average Approximation Ratios (AARs) of the best possible graph partition. The optimal solution of every problem instance is calculated by brute force and the AAR describes the deviation of given solution from it. An AAR of $1$ indicates perfect results, while an AAR of $0$ is the worst possible solution.

\begin{figure}[t]
  \centering
  \includegraphics[width=\columnwidth]{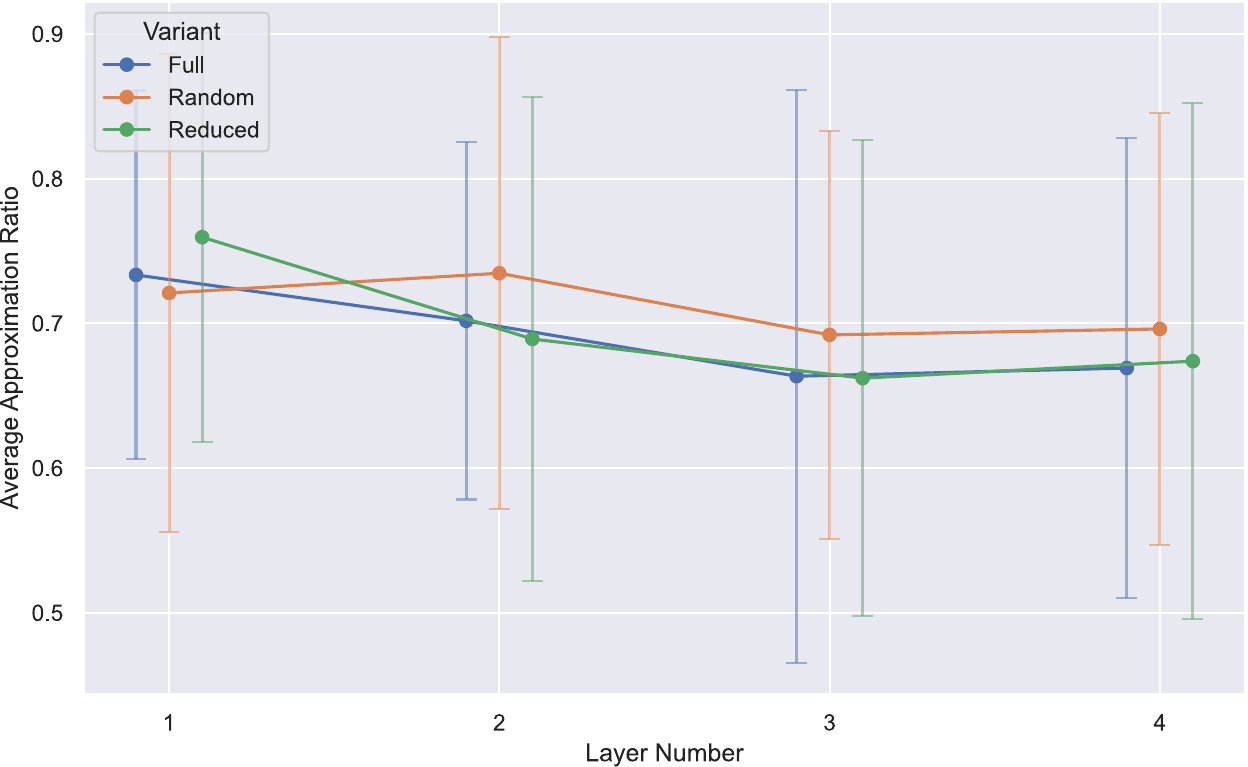}
  \caption{Average Approximation Ratio of the different circuit design schemes. Higher is better.}
  \label{fig:average_approx}
\end{figure}

Fig.~\ref{fig:average_approx} illustrates the evolution of the different AARs depending on the number of layers they implement. The plotted results show that the AARs generally decrease slightly with an increasing number of layers, but no clear systematic difference between the curves is apparent. In line with experience in QNN performance, further tests (not displayed here) without noise show the usually expected increase in performance for increasing circuit depth. Even though the noise level is quite small (merely a 1\% error), related work on very similar problems has shown that including circuit noise in training can severely deteriorate the overall performance comparable to our findings \cite{borras2023impact,10.1145/3489517.3530400,garcia2023effects}. While the random-width design shows a slightly better ARR performance in layer four compared to the the other two approaches, our small case study limits general conclusions on this advantage. Nevertheless, our data hints towards an advantage of the random pruning approach. 

Combining these results with the runtimes, we can clearly see, that the pruning based circuit designs train substantially faster while achieving the same solution quality in the presence of noise.

\section{\uppercase{Conclusion}}
\label{sec:conclusion}
In this work, we proposed the novel \emph{reduced-width} ansatz design pattern, that uses gate pruning to improve the trainability of a QNN. Our experiments have shown that this ansatz design can significantly speed training time (up to 5 times) on noisy quantum simulators while achieving similar or even better solution quality. Future work must verify its applicability to a broader range of base ansätze as well as its scaling behavior for larger circuits and problem instances. To explore beyond our noisy test environment, noise free simulations should be carried out, while also exploring different parameter training approaches. In conclusion, we presented a potent, novel ansatz design pattern for quantum machine learning, that opens up a new path towards trainable yet expressive quantum neural networks.
 


%

\section*{\uppercase{Acknowledgements}}
This paper was partially funded by the German Federal Ministry of Education and Research through the funding program "quantum technologies - from basic research to market" (contract number: 13N16196). Furthermore, this paper was also partially funded by the German Federal Ministry for Economic Affairs and Climate Action through the funding program "Quantum Computing -- Applications for the industry" (contract number: 01MQ22008A).

\bibliographystyle{apalike}
{\small
\bibliography{main}}


\end{document}